# Fractal Properties of Biophysical Models of Pericellular Brushes Can Be Used to Differentiate Between Cancerous and Normal Cervical Epithelial Cells

Juan de Dios Hernández Velázquez[1], Sergio Mejía-Rosales[1], and Armando Gama Goicochea[2*]

[1]Centro de Investigación en Ciencias Físico – Matemáticas (CICFIM), Universidad Autónoma de Nuevo León, Nuevo León 66450, Mexico

[2]División de Ingeniería Química y Bioquímica, Tecnológico de Estudios Superiores de Ecatepec, Estado de México 55210, Mexico

## ABSTRACT

Fractal behavior is found on the topographies of pericellular brushes on the surfaces of model healthy and cancerous cells, using dissipative particle dynamics models and simulations. The influence of brush composition, chain stiffness and solvent quality on the fractal dimension is studied in detail. Since fractal dimension alone cannot guarantee that the brushes possess fractal properties, their lacunarity was obtained also, which is a measure of the space filling capability of fractal objects. Soft polydisperse brushes are found to have larger fractal dimension than soft monodisperse ones, under poor solvent conditions, in agreement with recent experiments on dried cancerous and healthy human cervical epithelial cells. Additionally, we find that image resolution is critical for the accurate assessment of differences between images from different cells. The images of the brushes on healthy model cells are found to be more textured than those of brushes on model cancerous cells, as indicated by the larger lacunarity of the former. These findings are helpful to distinguish monofractal behavior from multifractality, which has been found to be useful to discriminate between immortal, cancerous and normal cells in recent experiments.

Keywords: cancer cells; pericellular brushes; polymer brushes; fractal dimension; lacunarity; dissipative particle dynamics

---

[*] Corresponding author. Electronic mail: agama@alumni.stanford.edu

# INTRODUCTION

The fractal concept introduced by Mandelbrot [1] is commonly invoked in the description of biological systems, where the repetition of patterns at different scales is frequent. In physiology, fractality has been used for decades [2, 3] in the analysis of complex patterns that can be found in neuronal and cardiac activity [4-7], arterial and blood vessel networks [8-11] and bronchial trees [12, 13], for example. Much of the physio – pathological research related to the fractal analysis of images has focused on correlating the fractal dimension (FD) of the structures or patterns present in such images with the health of cells. Among the extensive research in this field, there are works that relate the value of the FD to the presence of cancer [14-22]; studies carried out at the macro- and micro-scale for samples of colon [16], breast [17, 18], skin [19], cervical cells [20, 21], and even white blood cells [22] establish a difference between the FD of cancerous and normal tissues. From these results, it has been possible to distinguish healthy cells from cancerous cells [16, 17, 20, 22], define a relation with tumor growth [18, 23], trace the progression towards cancer [21] and measure its invasiveness [24].

In the context of the current understanding of the molecular mechanisms of cancer, medical imaging remains one of the most commonly used routes toward diagnosis. The implementation of fractal analysis for medical imaging has the potential of becoming a strong tool to yield precise diagnosis. The FD of an image can be estimated using several methods, such as box – counting, correlation, and Fourier analysis, among others [25]. Atomic force microscopy (AFM) has been widely used in cancer research to characterize mechanical properties that can discriminate between cancerous and normal cells [26-28]. Recently, it has been applied to the generation of topography and adhesion maps of individual cervical epithelial cells, wherein the FD of such images is calculated by Fourier analysis [20, 21]. The results show that the FD of cancerous cells tends to be higher than that of their healthy counterparts, with the differences likely due to the topographical features arising from the molecular brushes on the cell's surface. These brush – like structures coating the cell's surface are composed of complex macromolecules (microvilli, microtubules, microridges) tethered to the cell membrane, and results obtained with AFM show that their mechanical response can be measured separately from that of the cell's surface [26]. Furthermore, these experiments reveal that the brush on normal cervical cells (NCC) is made up of an approximately monodispersed array of chains, while cancerous cervical cells (CCC) are covered by a brush with at least two characteristic lengths. It is also argued that the grafting density on NCC brushes is lower than that on CCC brushes [26]. However, previous studies did not address aspects such as brush stiffness/softness, which is hypothesized to matter as cancer progresses [27], or the physicochemical environment of the cells, an important aspect to investigate since most experiments are carried out *in vitro*. Also, it is crucial to determine to what extent are the fractal properties of brushes dependent on image resolution, so that a confidence margin can be established when assessing distinctions between cancerous and normal cells.

**MODELS AND METHODS**

Here we report predictions of geometric properties of brushes that help distinguish between model NCC and CCC using numerical simulations. The models are solved using the dissipative particle dynamics (DPD) method [29, 30]. The CCC brush model is a tri-modal brush made up of end-grafted bead – spring linear chains of three different values of the polymerization degree: 294 short, 82 medium-sized and 33 large chains made of $N_1 = 5$, $N_2 = 30$ and $N_3 = 42$ DPD beads, with grafting densities of $\Gamma_1 = 1.74\ nm^{-2}$, $\Gamma_2 = 0.49\ nm^{-2}$ and $\Gamma_3 = 0.20\ nm^{-2}$, respectively. The NCC brush model is a monodispersed brush made up of 130 chains of $N = 27$ beads and a grafting density of $\Gamma = 0.78 \ nm^{-2}$. Brush models with these characteristics have been used previously [31] to reproduce accurately the mechanical response of brushes on human epithelial cervical cells under the AFM probe [26]. The novelty of the models introduced here is an added force between consecutive bonds, which controls the persistence length of the chains. By varying this three – body force one can control the local rigidity of the chains to define soft and stiff brushes. The motivation for considering brushes with different stiffness comes from tests carried out on mammary epithelial cells, where Young's modulus for cells at different tumorigenic phases is found to be lower as cancer progresses [27]. It is argued that the stiffness of the cells depends on their microenvironment. Here, the environment change is modeled as the change of the brush – solvent interactions, hence our brushes are modeled under good and bad solvent conditions. To complete the model, the brushes are confined by an explicitly curved surface made of DPD beads, to mimic their interaction with a nanosized AFM probe. Full details are provided in the Supplementary Information (SI).

Polymer brushes are created from polymer chains at relatively high grafting density; polymer chains are made up of linear sequences of monomeric beads, joined by freely rotating, harmonic springs:

$$\boldsymbol{F_S} = -k_s(r_{ij} - r_0)\hat{\boldsymbol{r}}_{ij}, \qquad (1)$$

where $k_s$ is the spring constant, and $r_0$ is the equilibrium position [32]. To model the chain's stiffness or softness, a three – body force acting between three consecutive beads is added [33]:

$$F_A = k_\theta \sin(\theta_{ijk} - \theta_0), \qquad (2)$$

where $k_\theta$ is the constant for the angular forces, [34, 35]. The relative distance between adjacent beads and the unit vector joining them are represented by the symbols $r_{ij}$ and $\hat{r}_{ij}$, respectively. The equilibrium angle is $\theta_0 = 180°$ and $\theta_{ijk}$ is the angle between two adjacent bonds, respectively. Appropriate values for the parameters in Eqs. (1) and (2) are chosen that prevent bond – crossing; for the Hookean spring the value $k_s = 100 \, k_B T / r_c^2$ [35] is used for all the chains in each system. For the angular force, two values of the constant $k_\theta$ are used to model the rigidity of the chains, namely $k_\theta = 10 \, k_B T / r_c$ for soft chains and $k_\theta = 100 \, k_B T / r_c$ for stiff chains. The systems are confined by two parallel surfaces perpendicular to the $z$–direction. The cell's surface ($z = 0$), is modeled by an effective, linearly decaying force given by:

$$\boldsymbol{F}_{wall} = \begin{cases} a_{iw}(1 - z_{iw}/z_c)\hat{\boldsymbol{z}} & z_{iw} \leq z_c \\ 0 & z_{iw} > z_c \end{cases}, \quad (3)$$

with a cutoff length $z_c$ and $z_{iw}$ being the distance between the $i$-th particle and the surface, $\hat{\boldsymbol{z}}$ is their unit vector, and $a_{iw}$ is the maximum intensity of the force [7 AGG 36]. On the opposite side of the simulation box is placed the surface of the AFM probe, which is a semi sphere made up of DPD particles frozen in space and curvature radius equal to $R = 0.8 \, L_x$. The non - bonding conservative DPD force is:

$$\boldsymbol{F}_{ij}^{C} = \begin{cases} a_{ij}(1 - r_{ij}/r_c)\hat{\boldsymbol{r}}_{ij} & r_{ij} \leq r_c \\ 0 & r_{ij} > r_c \end{cases}, \quad (4)$$

where $a_{ij}$ is the repulsion parameter between beads $i$ and $j$ and $r_c$ is the cutoff distance, set to $r_c = 1$. The former depends on the coarse –graining degree (the number of water molecules grouped into a DPD particle); for a coarse – graining degree equal to three, $a_{ii} = 78.0 \, k_B T / r_c$, where $k_B$ is Boltzmann's constant, and $T$ is the absolute temperature.

Table I shows the values of the interaction parameter for the various pairs of particles. Full additional details can be found in the SI.

**Table 1**. Interaction parameters $a_{ij}$ of the conservative force DPD and the force of the implicit surface representing the cell surface. $k_B T$ and $r_c$ are expressed in reduced DPD units and represent energy and length, respectively. Bead species are: S = solvent, H = chain's head, T = chain's tail, P = AFM probe and C = cell's surface.

| $a_{ij}[k_B T/r_c]$ j <br> i | S | H | T | P | C |
|---|---|---|---|---|---|
| S | 78 | 79.3(85) | 79.3(85) | 140 | 100 |
| H | 79.3(85) | 78 | 78 | 140 | 60 |
| T | 79.3(85) | 78 | 78 | 140 | 100 |
| P | 140 | 140 | 140 | 78 | 0* |
| C | 100 | 60 | 100 | 0* | 0** |

*Since the distance is larger than the cutoff radius. **Because the cell's surface is implicit.

The FD is calculated by Fourier analysis, following the procedure used to process images of human cervical epithelial cells [20, 21]. The procedure requires the calculation of the two–dimensional Fast Fourier Transform of the image as follows:

$$F(u,v) = \frac{\sum_{x=0}^{N_x-1} \sum_{y=0}^{N_y-1} z(x,y) \exp\left[-i2\pi\left(\frac{ux}{N_x}+\frac{vy}{N_y}\right)\right]}{N_x N_y}, \quad (5)$$

where $z(x, y)$ is the height of the brush at the pixel $(x, y)$, and $N_x$ and $N_y$ are the number of pixels in the $x$ and $y$ directions, respectively. Then, the magnitude of $F(u, v)$ is transformed into polar coordinates and averaged over the angle:

$$A(Q) = 1/\pi \int_0^\pi F(Q\cos\varphi, Q\sin\varphi)d\varphi. \quad (6)$$

$Q$ is the inverse of the lateral size $L$ of the geometrical features on the image. Linear behavior of $A(Q)$ on log-log scale ($A(Q) \sim Q^b$) is a signature of fractality. We extracted images with $L_x = L_y = 14\ nm$, which leads to $Q_{min} \approx 0.071\ nm^{-1}$. The FD, $\alpha$, as suggested by Dokukin et al. [20], is defined as $\alpha = 2 - b$, so that for flat surfaces $b = 0$ ($\alpha = 2$), while $b = -1$ ($\alpha = 3$) for infinitely rough surfaces, as limiting cases.

**RESULTS AND DISCUSSION**

The very concept of fractality implies that it is possible to find self – similar structural features on images upon magnifications on any scale. However, in practice, finding structures that are invariant over large orders of magnification is impossible; this empirical fact restricts the scale ranges where fractality can be observed. In image processing such ranges are generally restricted by the resolution of the image, since the search for patterns on scales smaller than the size of a pixel becomes meaningless. We calculate the FD for each image generated from DPD simulations at four resolutions: 1024×1024, 512×512, 256×256 and 128×128 pixels ($N_x \times N_y$). Figure 1 shows some representative examples of height images of the brushes, at a resolution of 1024×1024 pixels; in the SI we describe the procedure followed to generate the images.

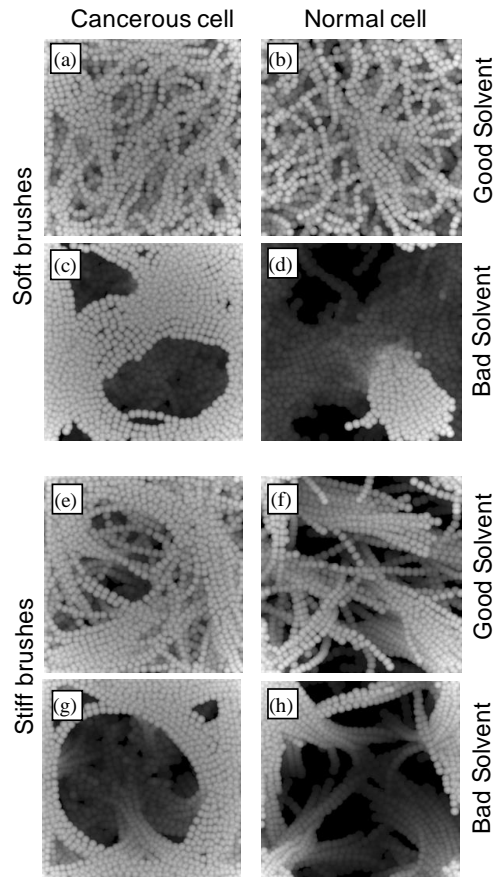

**FIG. 1**. Height images of the tri- and monodisperse brushes (cancer and normal cell brushes, respectively) under good solvent conditions (panels a, b, e and f), and under bad solvent condition (panels c, d, g and h). The resolution of the images is 1024×1024 pixels. The images are taken from the top of the simulation box and show the height in gray scale on the *XY* – plane, or frontal view. The beads are the monomers that make up the polymeric chains.

The FD obtained at different resolution for soft brushes is presented in Fig. 2(a), where it is noted that the FD of the polydisperse (CCC) brushes is higher than that of the monodisperse (NCC) brushes. These results are in agreement with recent experiments on cancer cells showing higher FD than their normal counterparts [20, 22], where the difference in the FD between CCC and NCC brushes becomes more significant when they are immersed in a bad solvent. Our predictions for soft brushes yield values of the FD in the range $2.2 < \alpha < 2.8$, in very good quantitative agreement with the work of Dokukin *et al.* [20], who find $2 < \alpha < 2.6$. The FD is restricted by the Euclidian dimensions of the systems [1]. For $\alpha = 2$, the brush should form an ideal flat surface while for $\alpha = 3$ all the chains that make up the brush should be fully stretched along the direction normal to the cell's surface and distributed over it in such a way that maximizes roughness.

As shown in Fig. 2(a), the FD of soft brushes on CCCs under good solvent conditions (Fig. 1(a)) is higher than that of the brush under bad solvent conditions (Fig. 1(c)); the same trend is observed for soft brushes on NCCs (Figs. 1(b) and 1(d)). From their respective height images in Fig. 1, one notices how the brushes in bad solvent tend to collapse towards the surface of the cell, yielding lower roughness. The fact that the FD of the CCC brushes is higher than that of the NCC brushes becomes more noticeable under bad solvent conditions because in good solvent the chains are distributed more homogeneously over the space between the cell surface and the tip of the AFM (as seen in Figs. 1(a) and 1(b)). By contrast, in poor solvent the end-grafted chains form mushroom-like structures and are displaced towards the walls of the system (in this case, the surface of the cell and the tip of the AFM) due to the relatively higher repulsive interactions between the solvent and the brush (see Figs. 1(c) and 1(d)).

The FD predicted for *stiff* brushes, in Fig. 2(b), displays contrasting trends with those for *soft* brushes (Fig. 2(a)). Under good solvent conditions, solid symbols in Fig. 2(b), the FD of stiff brushes on CCCs and NCCs is practically the same ($2.6 \lesssim \alpha \lesssim 2.65$), with small variations due to image resolution. This is the result of the three – body, angular interactions on stiff brushes dominating over the brush structure (monodispersed or polydispersed), see Figs. 1(e) and 1(f). Under good solvent conditions the non – bonding conservative DPD interactions between chains and solvent are smaller by definition than the angular interactions along the chains (see the SI), leading to images that are insensitive to the polymerization degree, as shown in Figs. 1(e) – 1(f). Consequently, the FD extracted from those images and shown by the solid symbols in Fig. 2(b) is insensitive to the monodispersity or lack thereof of the brushes, as is approximately the averaged FD for *soft* CCC and NCC brushes in good solvent, shown in Fig. 2(a). For stiff brushes under bad solvent conditions the FD of the polydispersed (CCC, Fig. 1(g)) brushes is once again larger than the FD of the monodispersed (NCC, Fig. 1(h)) brushes, as the open symbols in Fig. 2(b) show.

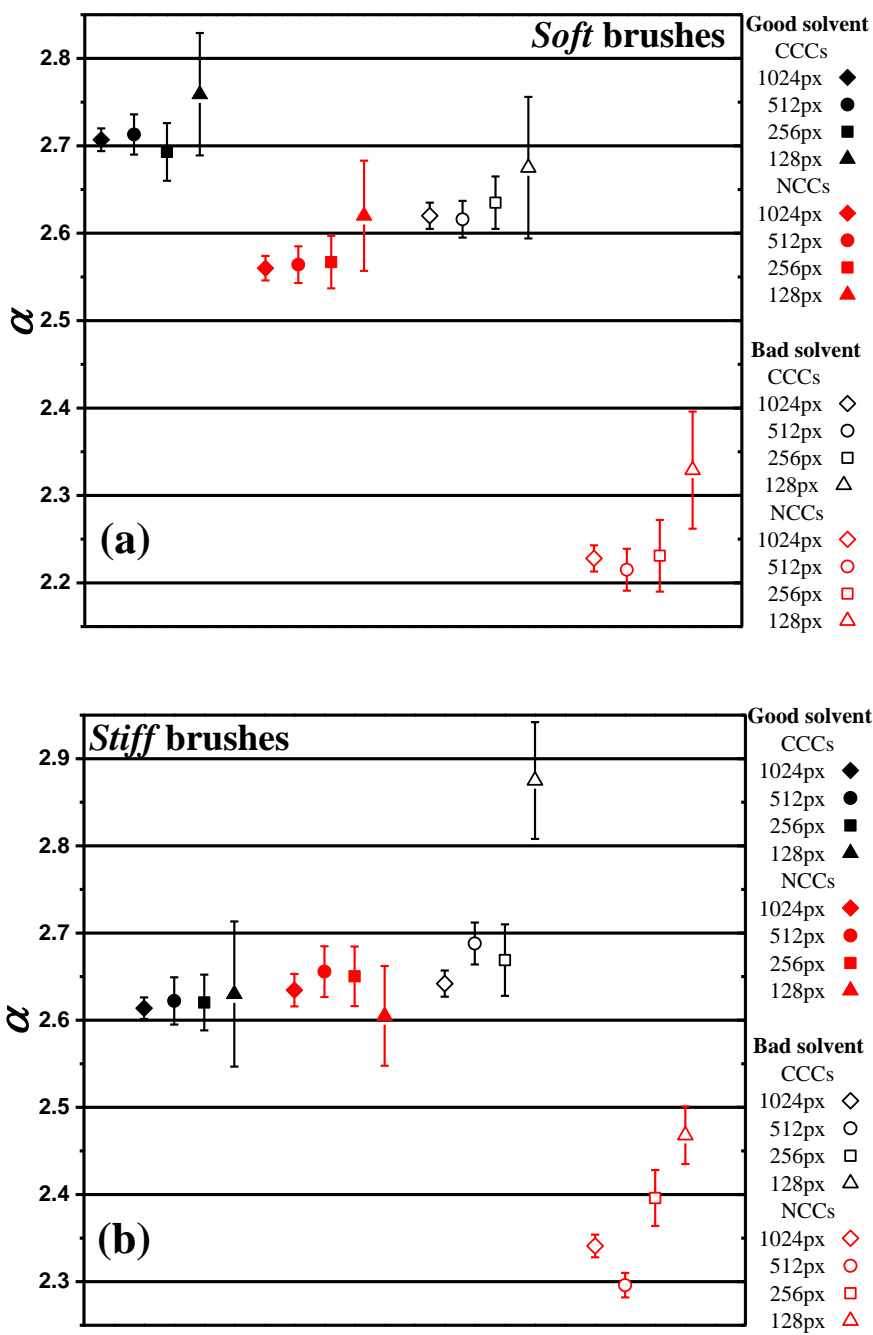

**FIG. 2**. (Color online) Fractal dimension of the height images (see Fig. 1) for *soft* (a) and *stiff* (b) brush models at different image quality. The error bars represent the standard error.

The results presented in Fig. 2 show that there exists a correlation between the standard error in the calculation of the FD and the image quality, and that the error is minimized as the resolution of the image is improved. This occurs because higher quality images display more features, which yield a broader range in which linear behavior (for $A(Q) \sim Q^b$ on log – log scale, see Eq. (6)) can be obtained from the features of the imaging. However, one can in principle obtain a FD from the image of an object which may not be fractal. Therefore, one needs additional tools to prove unequivocally the hypothesis that the structure of the brushes on the surface of cells can be used as a marker for diagnosis of their health condition. For that purpose, we have calculated the lacunarity of the brushes under both solvent conditions. Lacunarity ($\lambda$) is a concept introduced also by Mandelbrot [1] as a measure of the distribution of gaps or holes on an image. It gauges how much a geometric object departs from translational invariance or homogeneity [37]. For example, low lacunar objects are homogeneous and translationally invariant because they have gaps of about the same size or none at all, while high lacunar objects are heterogeneous and translationally covariant because of their wide distribution of gap sizes [38]. It is an additionally important variable to study because objects with the same FD can have very different lacunarity values [1], which provides additional information about the geometric properties of the objects. Lacunarity $\lambda(\varepsilon)$ is defined as the mean – square deviation of the variation of mass distribution probability $Q(M, \varepsilon)$ divided by its square mean:

$$\lambda(\varepsilon) = \frac{\sum_M M^2 Q(M,\varepsilon)}{[\sum_M M Q(M,\varepsilon)]^2}, \qquad (7)$$

where $\varepsilon$ is the box size and $M$ is the mass of the pixels or the mass of the grayscale image [39]. It is the pixels' mass density of an image, $\lambda(\varepsilon) = M(\varepsilon)/V(\varepsilon)$, hence if fractal geometry is present, then $M \sim \varepsilon^{FD}$, and the volume goes as $V \sim \varepsilon^{ED}$, where $ED$ is the Euclidean dimension. Therefore, fractality appears if a linear fit is obtained on a log $\lambda$ vs log $\varepsilon$ curve, with slope $m = FD - ED$. In Fig. 3, we report the values of the lacunarity of the images of CCC and NCC brush models at resolution of 1024×1024 pixels; full details can be found in the SI. It is to be remarked that this study of lacunarity should be considered as a proof of concept which could be implemented even in experimental images of pericellular brushes obtained through the AFM technique, as is shown in Fig. 4. A common feature found in both soft and stiff brushes (Fig. 3) is that the lacunarity of monodisperse brushes (NCC), at a given box size, under poor solvent conditions is larger than it is for all other cases, particularly for polydisperse brushes (CCC). For soft brushes in good solvent, see the solid symbols in Fig. 3(a), the differences between their lacunarities, whether they are mono – or polydisperse, are minimal. This is also apparent from the comparison of holes distributions in Figs. 1(a) and 1(b); when mono – and polydisperse brushes are soft and under good solvent they are equally lacunar. If they are stiff, Fig. 3(b) shows that brush composition and solvent quality affect strongly their lacunarity. Since lacunarity is a measure of the lack of translational invariance, our results show that monodisperse brushes on normal cells in bad solvent have more "gappiness" and are therefore less translationally and rotationally invariant than their polydisperse counterparts. This is a consequence of their different grafting densities, which may in turn be a consequence of the mechanisms that give rise to cell malignancy [40].

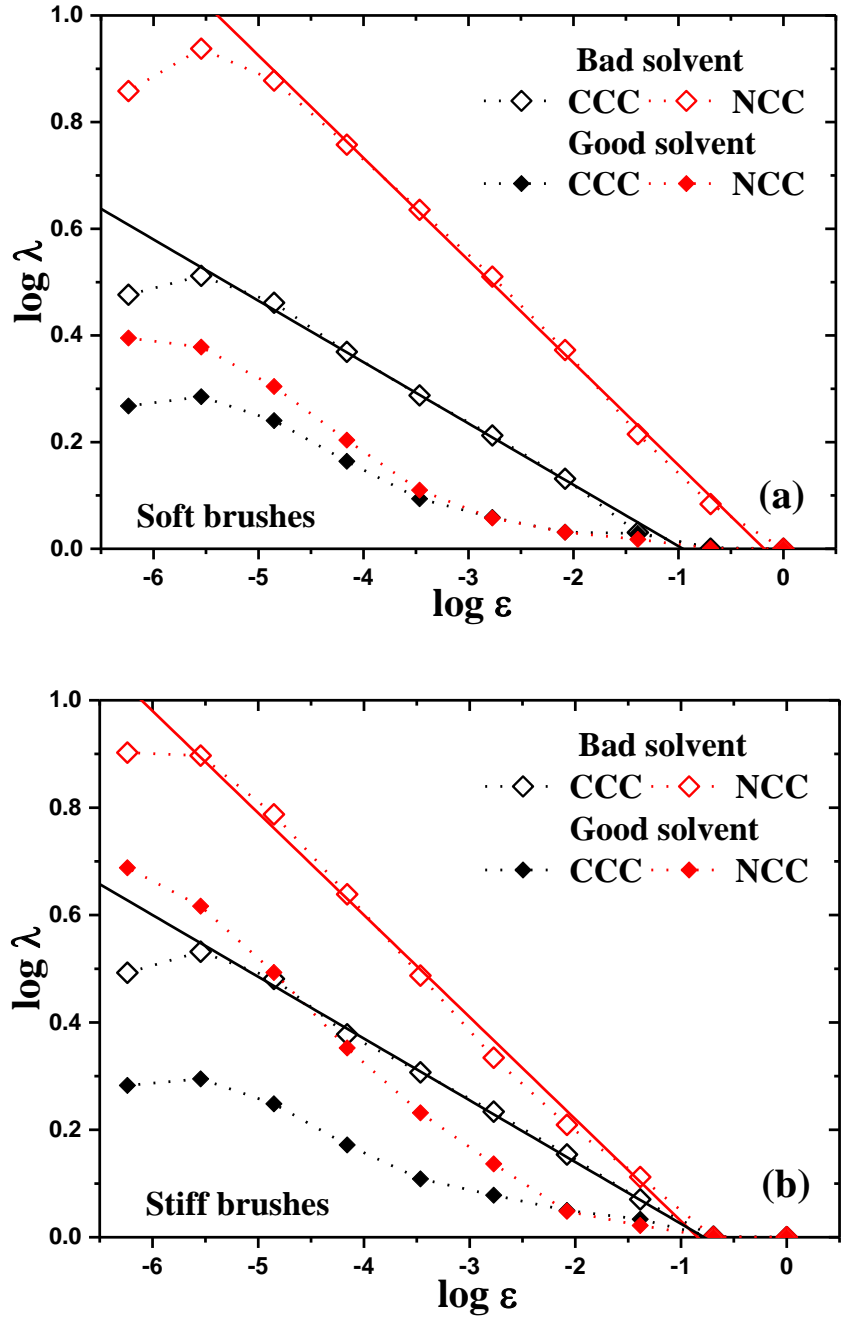

**FIG. 3**. (Color online) Lacunarity, λ, of images of *soft* (a) and *stiff* (b) brushes at resolution of 1024×1024 pixels, shown in Fig.1, as a function of box size, ε. The solid lines indicate fractal behavior. The software Image J [41] was used to obtain λ; full technical details are found in the SI.

Notice in Fig. 3 that the lacunarity of brushes under good solvent conditions cannot be approximated by a single linear fit, as two or more lines would be needed. This aspect indicates that the images of those brushes, in particular those on NCC, are not simply fractal and can only be considered as multifractal, in agreement with conclusions derived from experiments [20, 42].

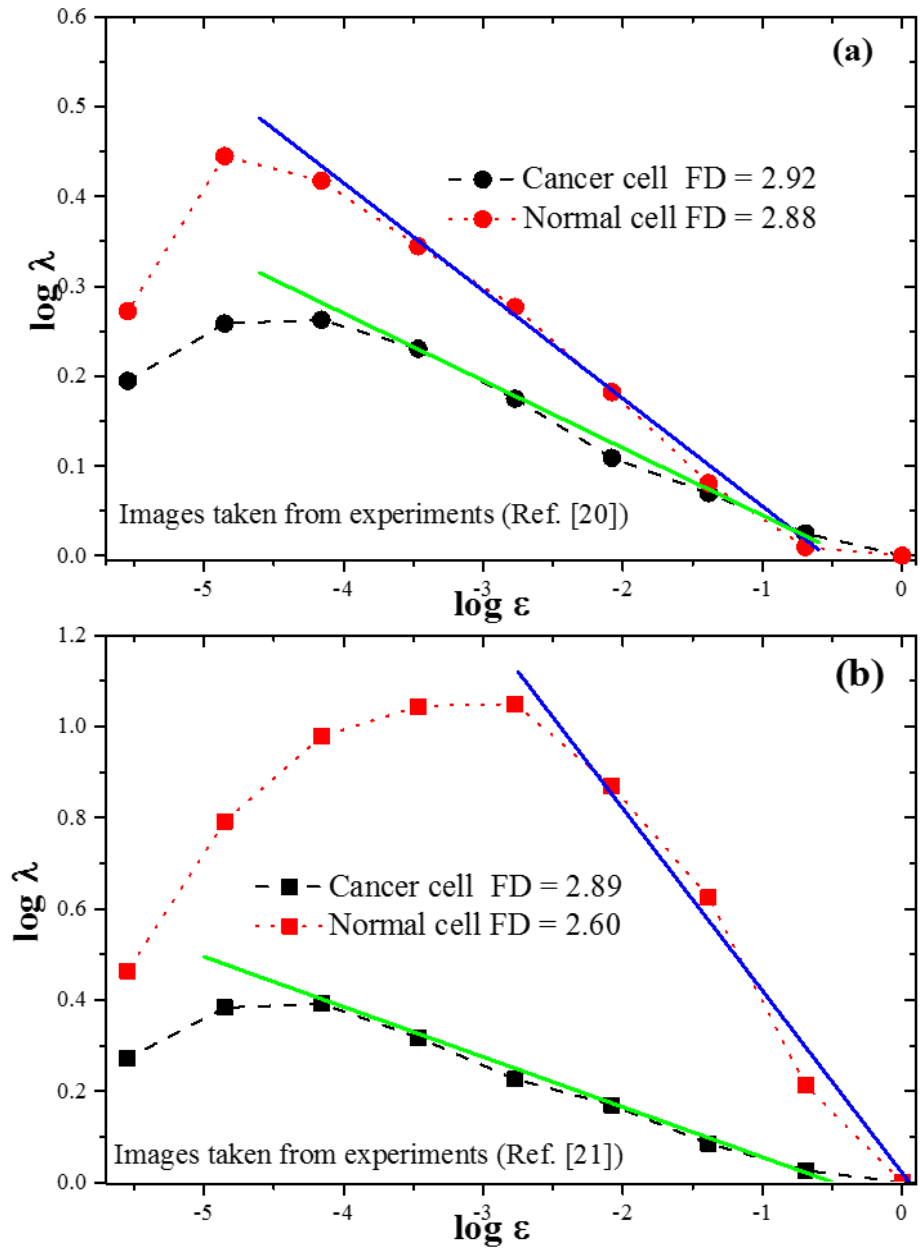

**FIG. 4**. (Color online) Lacunarity, λ, of images of brushes on cancerous and healthy human epithelial cells at resolution of 1024×1024 pixels, taken from experiments (reference [20] for Fig.4(a) and reference [21] for Fig. 4(b)), as a function of box size, ε. The solid lines indicate fractal behavior. The software Image J [41] was used to obtain λ;

full technical details are found in the SI. The legends report the fractal dimension obtained from the linear fit to the lacunarity, for each case.

In Figs. 4(a) and 4(b) there is a comparison of the lacunarity obtained from our analysis of images of brushes, taken from human cancerous and normal cervical epithelial cells using AFM [20, 21]. From the linear fits of the lacunarity one can obtain the brushes' FD, confirming that for brushes on cancerous cells FD is larger than that for brushes on normal cells, yielding differences of at least 10 % between cancerous and normal cells, in agreement with the difference reported in experiments [20, 21]. The calculation of $\lambda$ provides additional information that complements what one learns from the FD, because it helps discern from different types of patterns. Also, it yields size – dependent information, as opposed to single – value properties, such as the FD [43]. Additionally, analysis based on $\alpha$ and $\lambda$ has been found to be more sensitive in detecting apoptosis than cytometric assays [44]. We find in particular that soft brushes on normal cells (Fig. 1(c)) have larger $\lambda$ than soft brushes on cancer cells (Fig. 1(d)) under bad solvent conditions because, as those figures show, the image of the brush on a normal cell presents more gappiness than the image of the brush on a cancer cell, even though their FD have the *opposite* trend. Fractal dimension gauges the complexity of the brush (the more inhomogeneous the brush, the larger the FD) while $\lambda$ tracks the lack of rotational and translational invariance. Taken in conjunction, those two parameters ($\alpha$ and $\lambda$) can render a more robust diagnosis based on image analysis. For example, for images taken on fixed and freeze dried cells [20], where the environment acts as a bad solvent on the brushes, the analysis reported here can help to distinguish clearly cancer cells from normal ones because the difference between their FD can be up to 20 % (see empty symbols in Fig.2, and [20, 22]) but the difference in their $\lambda$ values at a given box size can be about 40 % larger (see Fig. 3(a)) or more (see solid squares, Fig. 4(b)).

## CONCLUSIONS

In conclusion, the FD of soft, polydisperse brushes is distinguishably larger than the FD of soft monodisperse ones, under poor solvent conditions. This is confirmed by the calculation of the lacunarity, which adds useful information about the homogeneity or lack thereof of the surfaces with brushes. Image resolution is found to be a key variable to improve the detection of fractal geometry. Lastly, soft brushes respond more noticeably to inhomogeneity induced by polydispersity than stiff ones, while good solvent quality tends to "wash out" the differences of fractal properties between poly – and monodisperse brushes. These conclusions are expected to be useful for other applications of polymer brushes, such as those in stimuli – responsive materials in solvents of varying quality [45].

## CONFLICTS OF INTERESTS

There are no conflicts of interest to declare.

## ACKNOWLEDGMENTS

The authors thank ABACUS, where most calculations were carried out, as well as Yoltla (UAM-I). JDHV thanks CONACYT for support and R. López – Rendón for help with computational resources. AGG acknowledges S. J. Alas, R. Catarino Centeno, D. R. Hidalgo – Olguín, E. Pérez, and I. Sokolov for enlightening conversations.

## REFERENCES

[1] Mandelbrot, B. B. *The Fractal Geometry of Nature*; WH Freeman and Company: New York 1983.

[2] West, B. J.; Goldberger, A. L. Physiology in Fractal Dimension. *Am. Sci*. **1987**, 75, 354-365.

[3] Goldberger A. L.; West, B. J. Fractals in Physiology and Medicine. *Yale J. Biol. Med.* **1987**, 60, 421-435.

[4] Esteller, R.; Vachtsevanos, G.; Echauz, J. Litt, B. A Comparison of Waveform Fractal Dimension Algorithms. *IEEE T. Circuits Syst.* **2001**, 48, 177-183.

[5] Liu, J. Z; Yang, Q.; Yao, B.; Brown, R. W.; Yue, G. H. Linear Correlation Between Fractal Dimension of EEG Signal and Handgrip Force. *Biol. Cybern.* **2005**, 93, 131-140.

[6] Acharya U., R.; Bhat, P. S.; Kannathal, N.; Rao, A.; Lim, C. M. Analysis of Cardiac Health Using Fractal Dimension and Wavelet Transformation. *ITBM-RBM* **2005**, 26, 133-139.

[7] Mishra, A. K.; Raghav, S. Local Fractal Dimension Based ECG Arrhythmia Classification. *Biomed. Signal Proces.* **2010**, 5, 114-123.

[8] Cross, S. S.; Start, R. D.; Silcocks, P. B.; Bull, A. D.; Cotton, D. W. K.; Underwood, J. C. E. Quantitation od the Renal Arterial Tree by Fractal Analysis. *J. Pathol.* **1993**, 170, 479-484.

[9] Beard, D. A.; Bassingthwaighte, J. B. The Fractal Nature of Myocardial Blood Flow Emerges from a Whole-Organ Model of Arterial Network. *J. Vasc. Res.* **2000**, 37, 282-296.

[10] Zamir, M. Fractal Dimensions and Multifractility in Vascular Branching. *J. Theor. Biol.* **2001**, 212, 183-190.

[11] Azemin, M. Z. C.; Kumar, D. K.; Wong, T. Y.; Wang, J. J.; Mitchell, P.; Kawasaki, R.; Wu, H. Age-Related Rarefaction in the Fractal Dimension of Retinal Vessel. *Neurobiol. Aging.* **2012**, 33, 194.e1-194.e4.

[12] Shlesinger, M. F.; West, B. J. Complex Fractal Dimension of the Bronchial Tree. *Phys. Rev. Lett.* **1991**, 67, 2106-2108.

[13] Schmidt, A.; Zidowitz, S.; Kriete, A.; Denhard, T.; Krass, S.; Peitgen, H-O. A Digital Reference Model of the Human Bronchial Tree. *Comput. Med. Imag. Grap.* **2004**, 28, 203-211.

[14] Sedivy. R. Chaodynamic Loss of Complexity and Self-Similarity in Cancer. *Med. Hypotheses* **1999**, 52, 271-274.

[15] Baish, J. W.; Jain, R. K. Fractals and Cancer. *Cancer Res.* **2000**, 6, 3683–3688.

[16] Esgair, A. N.; Naguib, R. N. G.; Sharif, B.S.; Bennett, M. K.; Murray, A. Fractal Analysis in the Detection of Colonic Cancer Images. *IEEE T. Inf. Technol. B.* **2002**, 6, 54-58.

[17] Rangayyan, R. M.; Nguyen T. M. Fractal Analysis of Contours of Breast Masses in Mammograms. *J. Digit. Imaging* **2007**, 20, 223-237.

[18] Tambasco M.; Magliocco, A. M. Relationship Between Tumor Grade and Computed Architectural Complexity in Breast Cancer Specimens. *Hum. Pathol.* **2008**, 39, 740-746.

[19] Timbó, C.; da Rosa, L. A. R.; Gonçalves, M.; Duarte, S. B. Computational Cancer Cells Identification by Fractal Dimension Analysis. *Comput. Phys. Commun.* **2009**, 180, 850-853.

[20] Dokukin, M. E.; Guz, N. V.; Gaikwad, R. M.; Woodworth, C. D.; Sokolov, I. Cell Surface as a Fractal: Normal and Cancerous Cervical Cells Demonstrate Different Fractal Behavior of Surface Adhesion Maps at the Nanoscale. *Phys. Rev. Lett.* **2011**, 107, 028101.

[21] Guz, N. V.; Dokukin, M. E.; Woodworth, C. D.; Cardin, A.; Sokolov, I. Towards Early Detection of Cervical Cancer: Fractal Dimension of AFM Images of Human

Cervical Epithelial Cells at Different Stages of Progression to Cancer. *Nanomed-Nanotechnol.* **2015**, 11, 1667-1675.

[22] Bauer, W.; Mackenzie, C. D. Cancer Detection on a Cell-by-Cell Basis Using a Fractal Dimension Analysis. *Acta Phys. Hung. N.S. Heavy Ion Physics* **2001**, 14, 43-50.

[23] Kikuchi, A.; Kozuma, S.; Sakamaki, K.; Saito, M.; Marumo, G.; Yesugi, T.; Taketani, Y. Fractal Tumor Growth of Ovarian Cancer: Sonographic Evaluation. *Gynecol. Oncol.* **2002**, 87, 295-302.

[24] Ahammer, H.; Helige, Ch.; Dohr, G.; Weiss-Fuchs, U.; Juch, H. Fractal Dimension of the Choriocarcinoma Cell Invasion Front. *Physica D* **2008**, 237, 446-453.

[25] Landini, G. Applications of Fractal Geometry in Pathology. In *Fractal Geometry in Biological Systems*; Iannaccone, P. M., Khokha, M., Eds.; CRC. Press: Florida, 1996; pp. 205 – 246.

[26] Iyer, S.; Gaikwad, R. M.; Subba-Rao, V.; Woodworth, C. D.; Sokolov, I. Atomic Force Microscopy Detects Differences in the Surface Brush of Normal and Cancerous Cells. *Nat. Nanotechnol.* **2009**, 4, 389-393.

[27] Guo, X.; Bonin, K.; Scarpinato, K.; Guthold, M. The Effect of Neighboring Cells on the Stiffness of Cancerous and Non-Cancerous Human Mammary Epithelial Cells. *New J. Phys.* **2014**, 16, 105002.

[28] Zhao, X.; Zhong, Y.; Ye, T.; Wang, D.; Mao, B: Discrimination Between Cervical Cancer Cells and Normal Cervical Cells Based on Longitudinal Elasticity Using Atomic Force Microscopy. *Nanoscale Res. Lett.* **2015**, 10, 482.

[29] Hoogerbrugge, P. J.; Koelman, J. M. V. A. Simulating Microscopic Hydrodynamic Phenomena with Dissipative Particle Dynamics. *Europhys. Lett.* **1992**, 19, 155-160.

[30] Español, P.; Warren, P. Statistical Mechanics of Dissipative Particle Dynamics. *Europhys. Lett.* **1995**, 4, 191-196.

[31] Gama Goicochea, A.; Alas Guardado, S. J. Computer Simulations of the Mechanical Response of Brushes on the Surface of Cancerous Epithelial Cells. *Sci. Rep.* **2015**, 5, 13218.

[32] Kremer, K.; Grest, G. S. Synamics of Entangled Linear Polymer Melts: A Molecular-Dynamics Simulation. *J. Chem. Phys.* **1990**, 92, 5057-5086.

[33] Shillcock, J. C.; Lipowsky, R. Equilibrium Structure and Lateral Stress Distribution of Amphiphilic Bilayers from Dissipative Particle Dynamics Simulations. *J. Chem. Phys.* **2002**, 10, 5048-5061.

[34] van Vliet, R. E.; Hoefsloot, H. C. J.; Iedema, P. D. Mesoscopic Simulation of Polymer–Solvent Phase Separation: Linear Chain Behaviour and Branching Effects. *Polymer* **2003**, 44, 1757-1763.

[35] Gama Goicochea, A.; Romero-Bastida, M.; López-Rendón, R. Dependence of Thermodynamic Properties of Model Systems on Some Dissipative Particle Dynamics Parameters. *Mol. Phys.* **2007**, 105, 2357-2381.

[36] Gama Goicochea, A. Adsorption and Disjoining Pressure Isotherms of Confined Polymers Using Dissipative Particle Dynamics. *Langmuir* **2007**, 23, 11656-11663.

[37] Gefen, Y.; Meir, Y.; Mandelbrot, B. B.; Aharony, A. Geometric Implementation of Hypercubic Lattices with Noninteger Dimensionality by Use of Low Lacunarity Fractal Lattices. *Phys. Rev. Lett.* **1983**, 50, 145-148.

[38] Plotnick, R. E.; Gardner, R. H.; O'Neill, R. V. Lacunarity Indices as Measures of Landscape Texture. *Landscape Ecol.* **1993**, 8, 201-211.

[39] Allain, C.; Cloitre, M. Characterizing the Lacunarity of Random and Deterministic Fractal Sets. *Phys. Rev. A* **1991**, 44, 3552-3558.

[40] Sokolv, I. Fractals: A Possible New Path to Diagnose and Cure Cancer? *Future Oncol.* **2015**, 11, 3049-3051.

[41] Schneider, C. A.; Rasband, W. S.; Eliceiri, K. W. NIH Image to ImageJ: 25 Years of Images Analysis. *Nat. Methods* **2012**, 9, 671-675.

[42] Dokukin, M.E., Guz, N.V., Woodworth, C.D., Sokolov I. Emergence of fractal geometry on the surface of human cervical epithelial cells during progression towards cancer. *New J. Phys.* **2015**, 17, 033019.

[43] Hidalgo-Olguín, D. R.; Cruz-Vázquez, R. O.; Alas-Guardado, S. J.; Domínguez-Ortiz, A. Lacunarity of Classical Site Percolation Spanning Clusters Built on Correlated Square Lattices. *Transp. Porous Med.* **2015**, 107, 717-729 .

[44] Pantic, I.; Harhaji-Trajkovic, L.; Pantovic, A.; Milosevic, N. T.; Trajkovic, V. Changes in Fractal Dimension and Lacunarity as Early Markers of UV-Induced Apoptosis. *J. Theor. Biol.* **2012**, 303, 87-92.

[45] LeMieux, M. C.; Usov, D.; Minko, S.; Stamm, M.; Tsukruk, V. V. Local Chain Organization of Switchable Binary Polymer Brushes in Selective Solvents. In *Polymer Brushes*; Advincula, R., Brittain, W. J., Caster, K. C., Rühe, J., Eds.; Weily: Berlin, 2004; pp. 427 – 440.

# Supplementary Information

## For

## Fractal Properties of Biophysical Models of Pericellular Brushes Can Be Used to Differentiate Between Cancerous and Normal Cervical Epithelial Cells

J. D. Hernández Velázquez[1], S. Mejía-Rosales[1], and A. Gama Goicochea[2]†

[1]Centro de Investigación en Ciencias Físico – Matemáticas (CICFIM), Universidad Autónoma de Nuevo León (UANL), Nuevo León 66450, Mexico

[2]División de Ingeniería Química y Bioquímica, Tecnológico de Estudios Superiores de Ecatepec (TESE), Estado de México 55210, Mexico

## COMPUTATIONAL DETAILS

The dissipative particle dynamics (DPD) model [1, 2] is used to design and simulate pericellular brushes models. In addition to the elemental forces required to model a fluid in DPD, additional conservative forces are required to model brushes on cervical cells and the confinement of the systems.

For the design of the chains that make up the brushes we use the bead – spring model with two bonding forces that represent the interactions between two and three consecutive beads of the chains. The first one is a two – body harmonic force that joins adjacent beads (eq. (S1)) [3]. The other is a three – body sinusoidal force acting between three consecutive beads to provide rigidity to the chains (eq. (S2)) [4]:

$$\boldsymbol{F_S} = -k_s(r_{ij} - r_0)\hat{\boldsymbol{r}}_{ij}, \quad \text{(S1)}$$

$$F_A = k_\theta \sin(\theta_{ijk} - \theta_0), \quad \text{(S2)}$$

where $k_s$ and $k_\theta$ are the constants for the Hookean and angular forces, respectively; $r_0 = 1/\sqrt[3]{\rho} \approx 0.7 r_c$ is the equilibrium distance between two adjacent beads when the global density is equal to three [5, 6]. The symbols $r_{ij}$ and $\hat{\boldsymbol{r}}_{ij}$ are the relative distance and the unit vector between two consecutive beads, respectively; the relaxation angle is $\theta_0 = 180°$ and $\theta_{ijk}$ is the angle between two adjacent bonds, respectively. We use $k_s = 100 \, k_B T / r_c^2$ [6] for all the chains in each system, and two values for the angular force constant $k_\theta$ to modify the rigidity of the chains, namely $k_\theta = 10 \, k_B T / r_c$ for *soft* chains and $k_\theta = 100 \, k_B T / r_c$ for *stiff* chains. These values of the constants preclude bond crossing between chains. In Fig. S1, we show a schematic representation of the systems designed and solved in this work.

---

†Corresponding author. Electronic mail: agama@alumni.stanford.edu

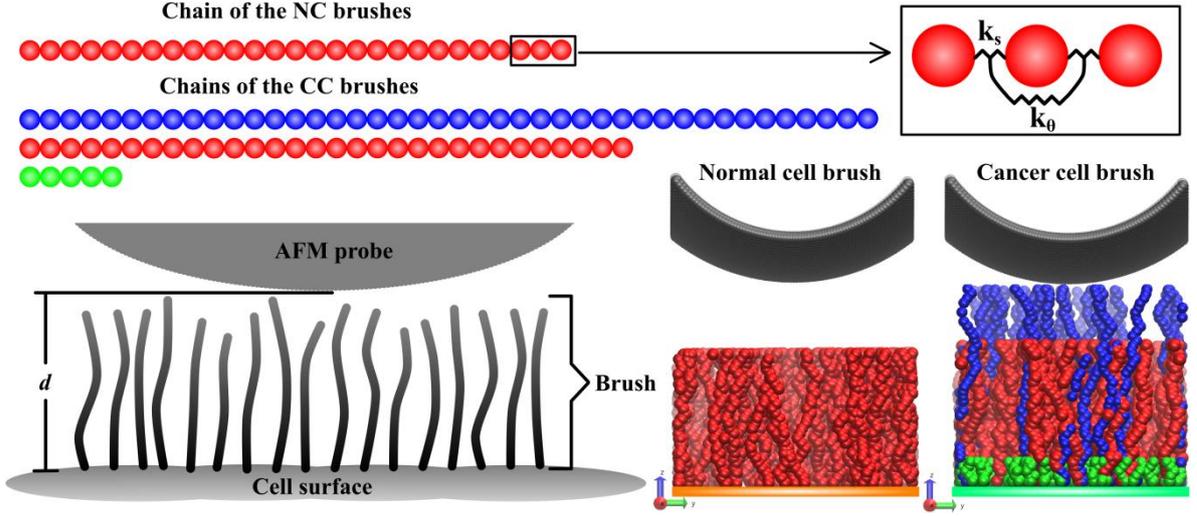

**Fig. S1**. Schematic overview of the models of pericellular brushes confined by an AFM probe used here. The solvent is omitted for clarity.

The systems are confined by two surfaces in the $z$–direction of the systems. The cell's surface, located at $z = 0$, is modeled through an effective square surface, wherein the interaction with the beads of the systems is given by:

$$\boldsymbol{F}_{wall} = \begin{cases} a_{iw}(1 - z_{iw}/z_c)\hat{\boldsymbol{z}} & z_{iw} \leq z_c \\ 0 & z_{iw} > z_c \end{cases}, \quad (S3)$$

which is a linearly decaying force with a cutoff length $z_c$; $z_{iw}$ is the distance between the i-th bead and the implicit surface, $\hat{\boldsymbol{z}}$ is their unit vector, and $a_{iw}$ is the maximum intensity of the force [7]. The surface of the AFM probe is included explicitly, i. e., with DPD particles; the curvature radius of the surface is $R = 0.8\,L_x$ and is located at $z = L_z = 26\,r_c$.

The quality of the solvent is varied to investigate the behavior of the brushes immersed in good and bad solvent conditions [8]. This is done by varying the interaction parameters for the conservative force DPD, which is more repulsive the larger the value of the parameter $a_{ij}$. The non - bonding conservative DPD force is:

$$\boldsymbol{F}_{ij}^C = \begin{cases} a_{ij}(1 - r_{ij}/r_c)\hat{\boldsymbol{r}}_{ij} & r_{ij} \leq r_c \\ 0 & r_{ij} > r_c \end{cases}, \quad (S4)$$

where $\boldsymbol{r}_{ij} = \boldsymbol{r}_i - \boldsymbol{r}_j$, $r_{ij} = |\boldsymbol{r}_{ij}|$ and $\hat{\boldsymbol{r}}_{ij} = \boldsymbol{r}_{ij}/r_{ij}$ are the relative position between particles $i$ and $j$ vector, its magnitude and its unit vector, respectively, and $a_{ij}$ is the repulsion parameter between two beads of different species. The latter depends on the coarse –graining degree, i. e., on the number of water molecules grouped into a DPD particle. For a coarse – graining degree equal to three, the interaction between the same bead species becomes $a_{ij} = 78.0\,k_B T/r_c$, where $k_B$ is Boltzmann's constant, $T$ is the absolute temperature of the system and $r_c$ is the cutoff radius, which is the reduced unit of the length in the DPD model and is commonly set to $r_c = 1$.

Table S1 lists the interaction parameters of the force representing the implicit cell surface (eq. (S3)) and the conservative force DPD (eq. (S4)), for each species of DPD particles.

**Table S1**. Interaction parameters $a_{ij}$ of the conservative force DPD and the force of the implicit surface representing the cell surface. $k_BT$ and $r_c$ are expressed in reduced DPD units and represent energy and length, respectively. Bead species are: S = solvent, H = chain's head, T = chain's tail, P = AFM probe and C = cell's surface.

| $a_{ij}[k_BT/r_c]$ j \ i | S | H | T | P | C |
|---|---|---|---|---|---|
| S | 78 | 79.3(85) | 79.3(85) | 140 | 100 |
| H | 79.3(85) | 78 | 78 | 140 | 60 |
| T | 79.3(85) | 78 | 78 | 140 | 100 |
| P | 140 | 140 | 140 | 78 | 0* |
| C | 100 | 60 | 100 | 0* | 0** |

*Since the distance is larger than the cutoff radius. **Because the cell's surface is implicit.

The dimensions of the simulation box are $L_x = L_y = 20\ r_c$ and $L_z = 26\ r_c$, where the value of $L_z$ is chosen so as to maintain a fixed distance $d = 10\ r_c$ between the cell surface and the tip surface of the AFM probe. All the simulations are carried out in two phases. The first one is the thermalization phase, which ends after 5 blocks of $10^4$ DPD time steps each, with $\Delta t = 0.01\ \tau$, which is sufficient for the system to reach thermal equilibrium. The second is the production phase, which consist of 20 blocks of $10^4$ DPD time steps of $\Delta t = 0.03\ \tau$, during which data are collected to obtain the average of the reported properties.

The reduced units used in this work are based on the coarse – graining ($N_m$) equal to three and a global density ($\rho^*$) equal to three, i.e., the total number of DPD particles divided by the available volume of the system. Length is reduced with the cutoff radius $r_c = 6.46$ Å [9]; the interaction parameters $a_{ij}$ and $a_{iw}$ are expressed in the reduced unit of force $k_BT/r_c = 1$; the latter parameters can change with temperature [10]. The time is reduced with $\tau = (mr_c^2/k_BT)^{1/2} \approx 3\ ps$, where $m \approx 9 \times 10^{-23}\ g$ is the mass of a water DPD particle with $N_m = 3$. Newton's second law of motion is integrated discreetly to solve the forces through a modified velocity Verlet algorithm [11, 12]. Periodic boundary conditions are used on the *x*– and *y*– directions of the simulation box since the system is confined in the *z* – direction.

**RESOLUTION OF IMAGES**

In this work, we use images at four different resolution values to determine which one is the best for the calculation of the fractal dimension (FD), considering that high – quality images require more sophisticated equipment and methodologies. The procedure followed to define the quality of the images is laid out below.

Step 1: The matrix of heights of the frame of the system on which we will work is extracted; in this case, the lateral size of the images is given in reduced DPD units, such that $L_x = L_y = 20\, r_c$.

Step 2: As the beads of the system technically are represented by points in space, the matrix of heights of the system is given by points on the *xy*–plane, where the intensity represents the height at which the bead is located. Thus, when the quality of the image $N_x \times N_y$ is chosen, one needs to save the positions $(p_x, p_y)$ in units of $r_c$ of each bead and convert that position to pixel coordinates $(x, y)$. Figure S2 shows a simple sketch of the process previously described.

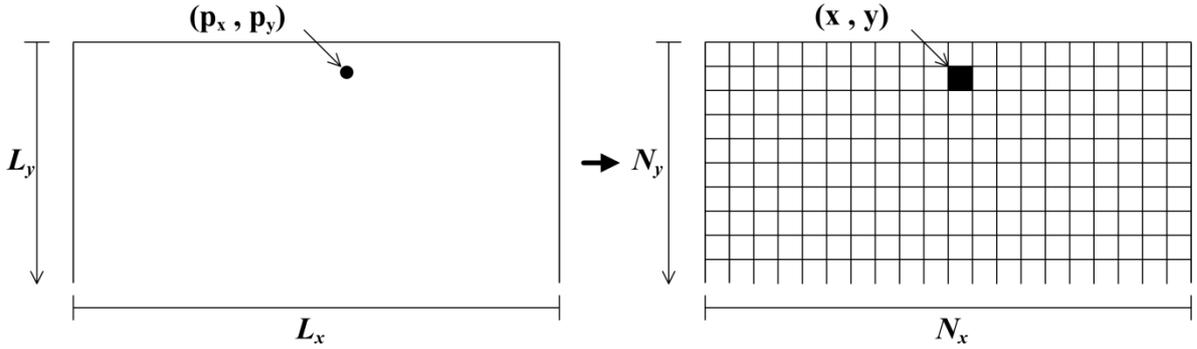

**Fig. S2**. Process used to assign coordinates for the points (beads) on the pixel map, where $L_x$ and $L_y$ are the lateral size of the system in units of $r_c$, while $N_x = N_y \in \mathbb{Z} > 0$, are the number of pixels in their respective directions.

The magnitude of one pixel in units of $r_c$ is given by

$$1 \text{ pixel} = L_x/N_x = L_y/N_y, \tag{S5}$$

where $L_x = L_y$ are given in units of $r_c$. For consistency, we choose values of $N_x = N_y$, which are the number of pixels in their respective directions. Hence, if the position for a point (bead) on the *xy*– plane is given by $(p_x, p_y)$, the location of such point (bead) on the pixel map is given by $\left(x = p_x * (N_x/L_x),\ y = p_y * (N_y/L_y)\right)$, rounding it up to the nearest positive integer to the right.

Step 3: Now that each point (bead) in the system has coordinates corresponding to the pixel map, one can "shape" every point. Since it is known that $r_c$ is the size of a DPD particle, its size in pixels can be found knowing that $1 r_c = N_x/L_x$ pixles; then, we fill the pixels with the value of the height given by

$$h(r) = z(x, y) + [r_c^2 - r^2]^{1/2}, \tag{S6}$$

where $z(x, y)$ is the value of the height for the pixel $(x, y)$. Figure S3 shows a schematic representation of equation (S6).

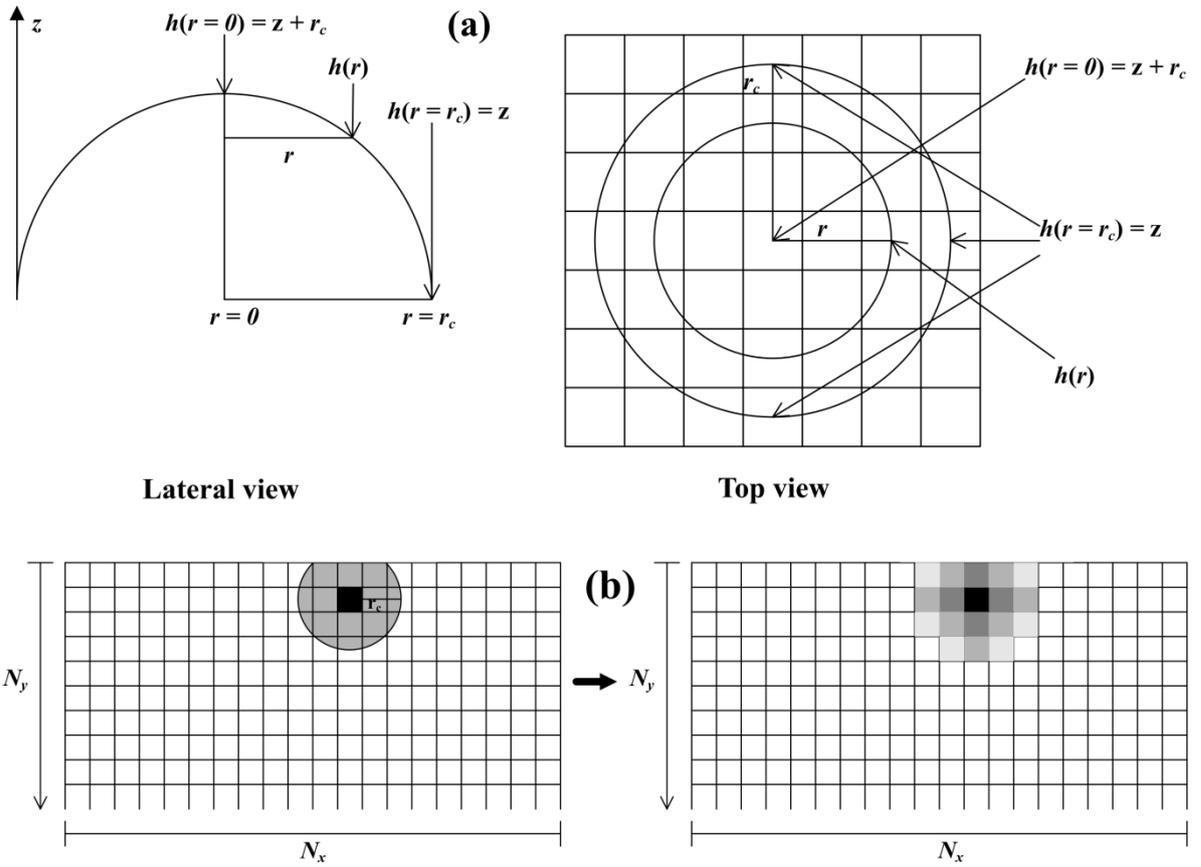

**Fig. S3**. Schematic representation of the height as a function of the distance $r$, with $r \in [0, r_c]$. (a) A sketch of the occupancy of a simple bead on the pixel map (b), where the intensity of the filled pixels represents the magnitude of the height.

In Fig. S4, we show the images at the four different resolutions for soft brushes covering a cancerous cell under bad solvent conditions, following the procedure just outlined.

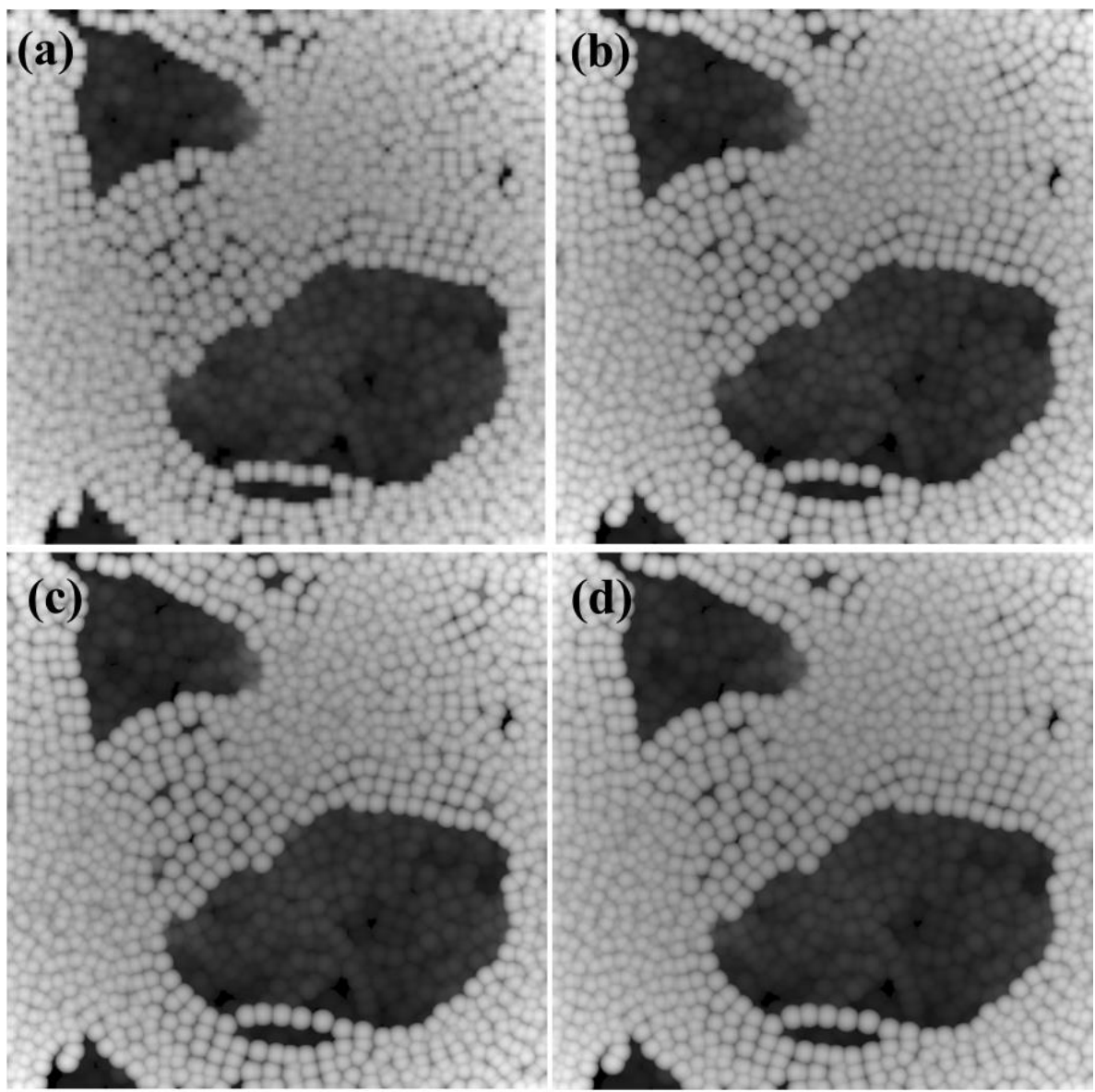

**Fig. S4**. Images of soft brushes on cancerous cell models under bad solvent conditions, at different resolution: 128×128, 256×256, 512×512 and 1024×1024, panels from (a) through (d), respectively. White (maximum intensity) represents the highest value of the brush height, while black (minimum intensity) represents the cell's surface height ($h = 0$).

**CALCULATION OF THE FRACTAL DIMENSION $\alpha$**

The FD is calculated from the 2 – dimensional Fast Fourier Transformed (FFT) images of the brushes. Figure S5 shows an example of the FFT image used in this work with their respective curve $A(Q)$ in log – log scale, where the slope $b$ of the fitting curve $log(y) = b\,log(Q) + log(a)$ is used to find the FD ($\alpha$) of the image through equation $\alpha = 2 - b$. The fitting was done in a range of $Q$ values within the range of the inverse size of the coarse – graining degree. This procedure is performed for all the images at their different resolutions (128×128, 256×256, 512×512 and 1024×1024 pixels).

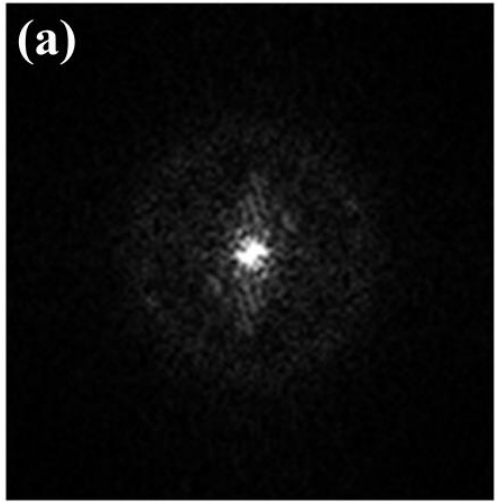

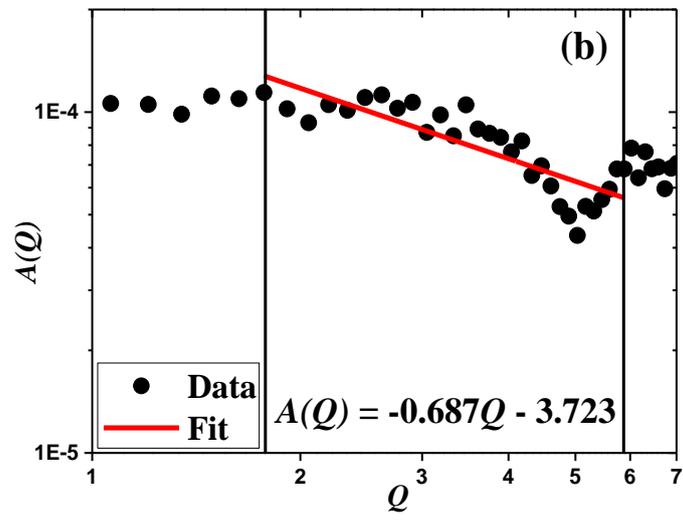

$A(Q) = -0.687Q - 3.723$

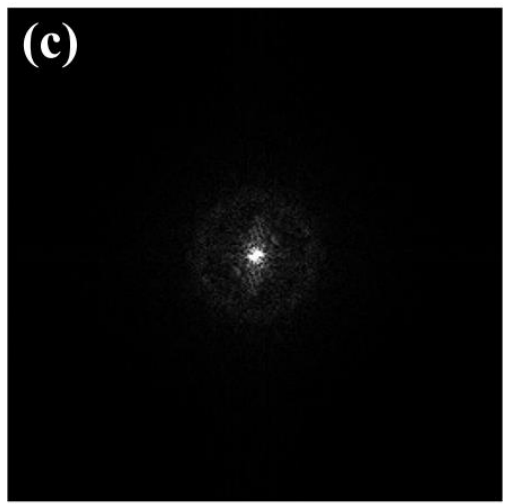

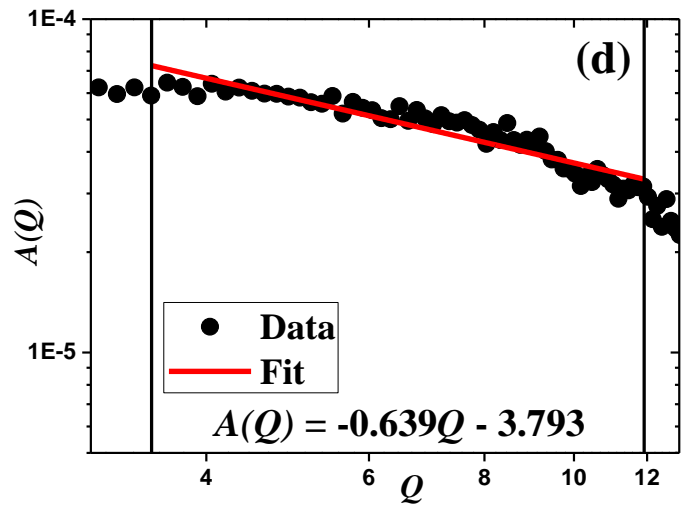

(d) $A(Q) = -0.639Q - 3.793$

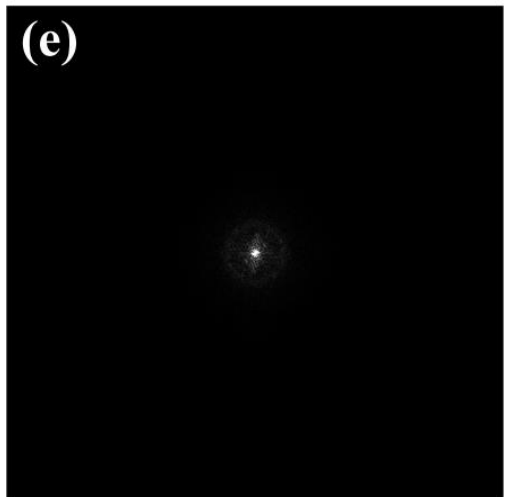

(e)

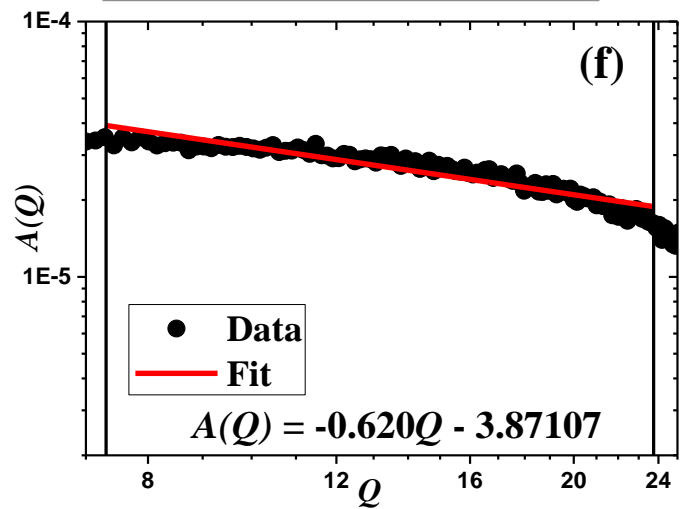

(f) $A(Q) = -0.620Q - 3.87107$

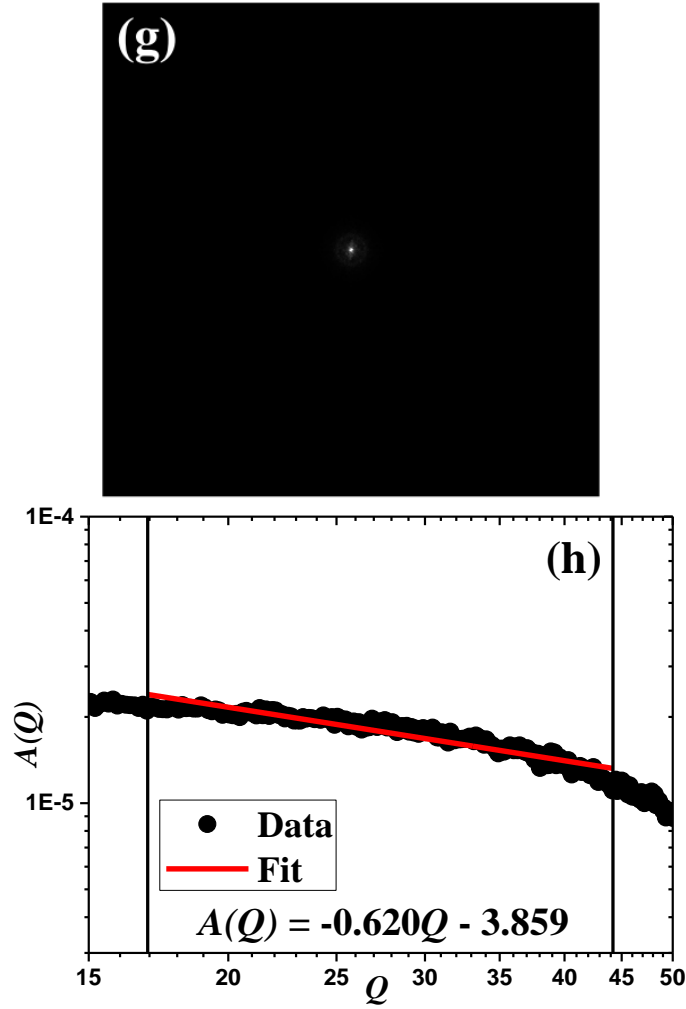

**Fig. S5.** Images of the 2 – dimensional FFT for soft brushes of model cancerous cells under bad solvent conditions with image quality of 128×128, 256×256, 512×512 and 1024×1024 pixels, panels (a), (c), (e) and (g), respectively. Panels (b), (d), (f) and (h) are their corresponding curves obtained from eq. 2 in the main article.

## CALCULATION OF THE LACUNARITY $\lambda$

Lacunarity was calculated with the plugin FracLac for ImageJ [13, 14], using the Differential Box Counting algorithm (DBC) [15]. This algorithm consists of placing a gliding-box of size $\varepsilon$ at a corner of an image window of size W×W, where $\varepsilon < $ W. Then, in the $\varepsilon \times \varepsilon$ gliding-box, a column made of $\varepsilon \times \varepsilon \times \varepsilon$ cubes is stacked to cover the maximum pixel value, where the minimum and maximum pixel values within the column are inside the cubes $u$ and $v$, respectively. Hence, the relative height of the column is given by [16, 17]

$$I(i, j, \varepsilon) = v - u + 1, \qquad (S7)$$

where $i$ and $j$ refer to pixel number of the image window. Once the gliding-box slides over the W×W section the mass of the grayscale image will be:

$$M(\varepsilon) = \sum_{i,j} I(i, j, \varepsilon), \qquad (S8)$$

thus, the lacunarity $\lambda$ can be obtained using eq. (3) in the main manuscript. In Fig. S6, we show an illustration of the main idea behind the DBC algorithm.

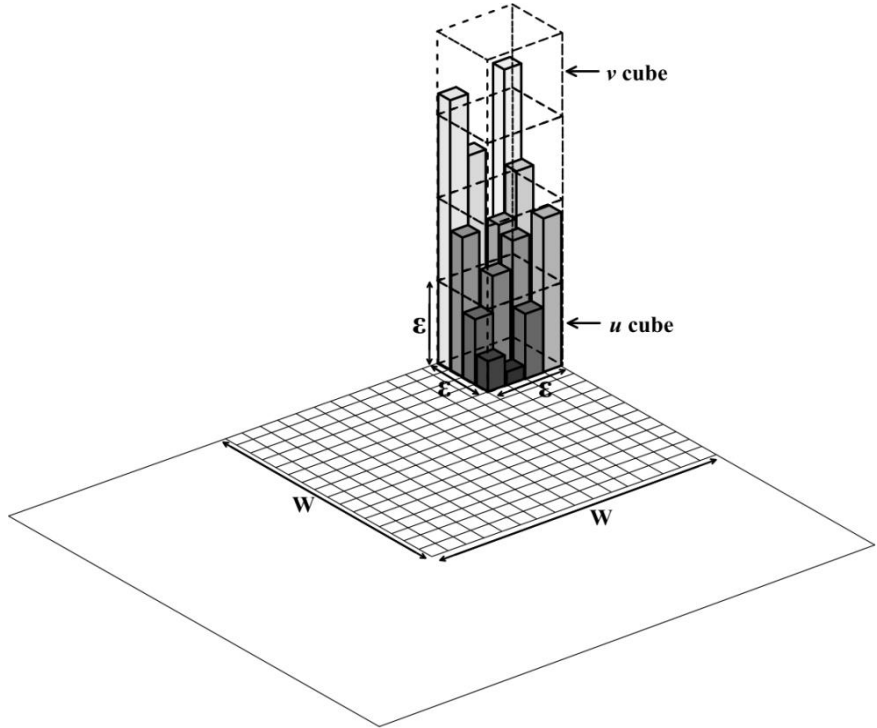

**Fig. S6**. Illustration of the DBC algorithm. In this example, for the 4×4 gliding box $u = 1$ and $v = 4$, therefore, $I = 4 - 1 + 1 = 4$.

We use a power series to scale the box size $\varepsilon$, with the base equal to 2 and exponent equal to 1; therefore, the minimum box size is 2×2 pixels. The maximum box size chosen was the size of the entire image. Since the resolution of the images is given by powers of 2, the maximum box size will depend on it. We also extract the FD of the images from these calculations, using the slope $m$ of the curve $\log \lambda(\varepsilon)$ vs $\log \varepsilon$, such that $m = D_f - D_e$, where $D_f$ is the FD and $D_e$ is the euclidean dimension [18]. Figure S7 shows the FD of the brushes images at the different resolutions, obtained from the lacunarity. Although the values of the FD obtained from the lacunarity curves differ from those obtained from the Fourier analysis, it is found that the trends are consistent with each other inasmuch as the CCC brushes images exhibit higher FD than the NCC brushes images and poor solvent condition enhance the fractality of the systems.

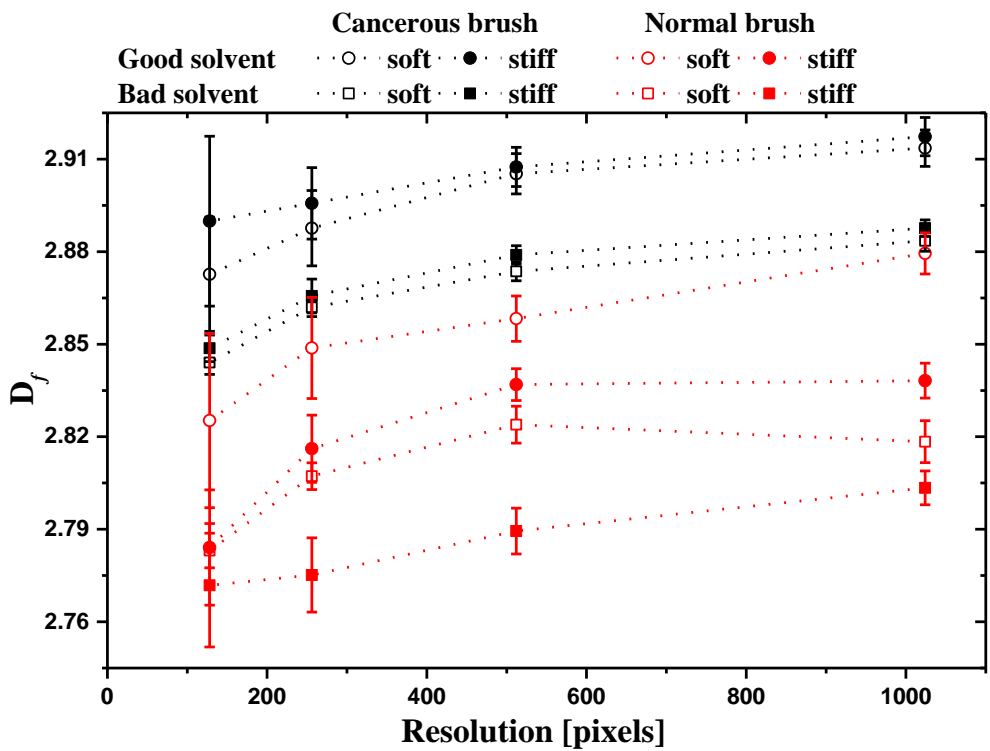

**Fig. S7**. Fractal dimension extracted from the slope of the log – log curve $\lambda$ vs $\varepsilon$, as function of the image quality. The error bars represent the standard error.

# REFERENCES

[1] Hoogerbrugge, P. J.; Koelman, J. M. V. A. Simulating Microscopic Hydrodynamic Phenomena with Dissipative Particle Dynamics. *Europhys. Lett.* **1992**, 19, 155-160.

[2] Español, P.; Warren, P. Statistical Mechanics of Dissipative Particle Dynamics. *Europhys. Lett.* **1995**, 4, 191-196..

[3] Kremer, K.; Grest, G. S. Synamics of Entangled Linear Polymer Melts: A Molecular-Dynamics Simulation. *J. Chem. Phys.* **1990**, 92, 5057-5086.

[4] Shillcock, J. C.; Lipowsky, R. Equilibrium Structure and Lateral Stress Distribution of Amphiphilic Bilayers from Dissipative Particle Dynamics Simulations. *J. Chem. Phys.* **2002**, 10, 5048-5061.

[5] van Vliet, R. E.; Hoefsloot, H. C. J.; Iedema, P. D. Mesoscopic Simulation of Polymer–Solvent Phase Separation: Linear Chain Behaviour and Branching Effects. *Polymer* **2003**, 44, 1757-1763.

[6] Gama Goicochea, A.; Romero-Bastida, M.; López-Rendón, R. Dependence of Thermodynamic Properties of Model Systems on Some Dissipative Particle Dynamics Parameters. *Mol. Phys.* **2007**, 105, 2357-2381.

[7] Gama Goicochea, A. Adsorption and Disjoining Pressure Isotherms of Confined Polymers Using Dissipative Particle Dynamics. *Langmuir* **2007**, 23, 11656-11663.

[8] Gama Goicochea, A.; Briseño, M. Application of Molecular Dynamics Computer Simulations to Evaluate Polymer–Solvent Interactions. *J. Coat. Technol. Res.* **2012**, 9, 279-286.

[9] Maiti, A.; McGrother, S. Bead–bead Interaction Parameters in Dissipative Particle Dynamics: Relation to Bead-Size, Solubility Parameter, and Surface Tension. *J. Chem. Phys.* **2004**, 120, 1594-1601.

[10] Mayoral, E.; Gama Goicochea, A. Modeling the Temperature Dependent Interfacial Tension Between Organic Solvents and Water Using Dissipative Particle Dynamics. *J. Chem. Phys.* **2013**, 138, 094703.

[11] Groot, R. D.; Warren, P. B. Dissipative Particle Dynamics: Bridging the Gap Between Atomistic and Mesoscopic Simulation. *J. Chem. Phys.* **1997**, 107, 4435-4435.

[12] Vattulainen, I.; Karttunen, M.; Besold, G. Integration Schemes for Dissipative Particle Dynamics Simulations: From Softly Interacting Systems Towards Hybrid Models. *J. Chem. Phys.* **2002**, 116, 3967-3979.

[13] Schneider, C. A.; Rasband, W. S.; Eliceiri, K. W. NIH Image to ImageJ: 25 Years of Images Analysis. *Nat. Methods* **2012**, 9, 671-675.

[14] Karperien, A. FracLac for ImageJ, http://rsb. info. nih. gov/ij/plugins/fraclac/FLHelp/Introduction. htm, **1999**-**2013**.

[15] Sarkar, N; Chaudhuri, B. B. An Efficient Approach to Estimate Fractal Dimension of Textural Images. *Pattern Recogn.* **1992**, 25, 1035-1041.

[16] Dong, P. Test of a New Lacunarity Estimation Method for Image Texture Analysis. *Int. J. Remote Sens.* **2000**, 21, 3369-3373.

[17] Barros Filho, M. N.; Sobreria, F. J. A. Accuracy of Lacunarity Algorithms in Texture Classification of High Spatial Resolution Images from Urban Areas. *Int. Arch. Photogramm.* **2008**, 37, 417-422.

[18] Stanley, H. E. Application of Fractal Concepts to Polymer Statistics and to Anomalous Transport in Randomly Porous Media. *J. Stat. Phys.* **1984**, 36, 843-860.